\documentclass[aip,twocolumn,letterpaper]{revtex4}

\usepackage{fullpage}

\usepackage{graphics}
\usepackage{epsfig}

\usepackage{xcolor}
\definecolor{ultramarine}{rgb}{0.07, 0.04, 0.56}
\usepackage{bm}
\usepackage{soul}
\usepackage[margin=1.0in]{geometry}

\begin{document}

\title{Magnetic reconnection and dynamos in the presence of plasma turbulence}
\author{Allen H Boozer}
\affiliation{Columbia University, New York, NY  10027 \linebreak ahb17@columbia.edu}

\begin{abstract} 

Evolving magnetic fields are frequently embedded in plasmas that are turbulent.  When the primary interest is in effects that are on a large scale compared to that of the turbulence, it is desirable to average over the turbulence to obtain equations for mean-field magnetohydrodynamics.   An obvious constraint on the validity of the averaging is that large-scale quantities that evolve slowly using the exact evolution equations must remain slowly evolving in the mean-field theory.  Magnetic helicity is the primary example of such a quantity and maintaining its slow evolution has \color{black} been controversial \color{black} in mean-field magnetohydrodynamics. A full theory of magnetic reconnection in turbulent plasmas is not the intent of this paper.   What is the intent is to show how exact results from Maxwell's equations explain why fast reconnection is so ubiquitous and what constraints these results place on the theory of magnetic field evolution, including dynamos, whether the plasma is turbulent or not.  These constraints are commonly broken in the reconnection literature, which has been heavily influenced by two-dimensional theory that is not applicable to three-dimensional problems.

  \end{abstract}

\date{\today} 
\maketitle


\section{Introduction}

A white paper written in 2020 by 108 members of the world reconnection community listed nine challenges to our understanding of magnetic reconnection \cite{Challenges:2020}.  In the same year, seven of those members wrote a review on plasma turbulence \cite{Turb-recon:2020} as a way of addressing reconnection challenges.  As shown by Boozer \cite{Boozer:rec2023} in 2023, the reconnection challenges can in large part be addressed by rectifying fundamental assumptions---many of which came from two-dimensional theories.  This paper has two purposes:  more complete explanations  (1) of the physics basis of required changes in assumptions and (2) of the constraints on the theory of reconnection in turbulent plasmas.

The validity of Maxwell's equations is fundamental to our knowledge of physics.   Any result derivable from Maxwell's equations provides a constraint on the validity of any magnetohydrodynamics simulation or mean-field theory---not the other way around.  Five results that are derivable from Maxwell's equations and simple mathematics provide particularly important constraints. 

\subsection{Results from Maxwell's equations \label{sec:Maxwell results}}

(1)  Section \ref{sec:B-ev} derives the exact electric field representation $\vec{E}+\vec{u}_\bot\times\vec{B}=-\vec{\nabla}\Phi + \mathcal{E}\vec{\nabla}\ell$.  This implies that Faraday's Law for $\partial \vec{B}/\partial t$ can always be written in the advection-diffusion form. The component of electric field along the magnetic field is $E_{||}=-\partial\Phi/\partial \ell + \mathcal{E}$ where $\ell$ is the distance along a field line.  $\mathcal{E}$ is required to satisfy boundary or periodicity conditions on $E_{||}$ and can be taken to be a field line constant.  The advective velocity $\vec{u}_\bot$ is the velocity of the magnetic field lines when $\mathcal{E}=0$.  A non-zero $\mathcal{E}$ is required to break magnetic field line connections.  

Forty years ago, Hassan Aref \cite{Aref,Aref:PF} showed that advection-diffusion equations have a special property, which also holds for magnetic fields.  When the velocity $\vec{u}_\bot$ is chaotic, then by the definition of chaos each streamline has neighboring lines that separate from it exponentially with an e-folding time $\tau_{ef}$.  With a chaotic $\vec{u}_\bot$, magnetic field line connections break within a timescale approximately an order of magnitude longer than $\tau_{ef}$ for any credible \color{black} three dimensional \color{black} magnetic evolution problem.  

(2) Section \ref{sec:helicity-ev} derives the evolution equation for magnetic helicity, $K=\int \vec{A}\cdot\vec{B} d^3x$, in a bounded volume of space when the flux of helicity across the bounding surface is zero.  $\vec{A}$ is the vector potential; $\vec{B}=\vec{\nabla}\times\vec{A}$.  The rate of change of the helicity in the volume is $-2\int\vec{E}\cdot\vec{B}\color{black} d^3x\color{black}$, or equivalently $-2\int \mathcal{E} \vec{B}\cdot\vec{\nabla}\ell d^3x$.  

The magnetic helicity is not only conserved in the absence of magnetic field line breaking but also in the presence of arbitrarily rapid magnetic field line breaking when the volume average of $\vec{E}\cdot\vec{B}$ is zero.  Using the traditional Ohms Law, in which $E_{||}=\eta j_{||}$, a concentration of $j_{||}$ into sheets of intense current density leads to a rapid breaking of field line connections but has essentially no effect on the rate of helicity evolution.   

As shown by Woltjer \cite{Woltjer:1958}, helicity conservation limits the fraction of the electromagnetic magnetic energy that can be quickly transferred to a plasma.  This somewhat surprising limitation is an observed feature of tokamak disruptions, and explains why the plasma current generally increases when the magnetic surfaces are suddenly destroyed in a fast magnetic reconnection \cite{Nardon:2023}.

(3) Section \ref{sec:energy-ev} discusses the evolution equation for the electromagnetic energy in a bounded volume of space when the flux of this energy across the bounding surface is zero.   When the flow speed of the plasma is non-relativistic, the magnetic energy density, \color{black} $B^2/2\mu_0$, \color{black}  is generally far larger than the electric energy density,  \color{black} $\epsilon_0 E^2/2$, \color{black} and the change in the electromagnetic energy accurately gives the change in the magnetic energy.  

The rate of change of the electromagnetic energy in the volume $-\int\vec{j}\cdot\vec{E} d^3x$ is the sum of two terms: an ideal exchange $-\int(\vec{u}_\bot\cdot\vec{f}_L) d^3x$, with the Lorentz force $\vec{f}_L\equiv \vec{j}\times\vec{B}$, and an exchange associated with magnetic connection breaking $-\int \mathcal{E} \vec{j}\cdot\vec{\nabla}\ell d^3x$.  Unlike helicity, the energy transfer can be made arbitrarily rapid by concentrating $j_{||}$ into thin sheets.  A concentration of $j_{||}$ into thin sheets need not enhance the ideal energy transfer since $j_{||}$ has no direct effect on the Lorentz force. 

 Two definitions of a fast magnetic reconnection are (i) rapid changes in the connections of magnetic field lines and (ii) a fast transfer of energy from the electromagnetic field into plasma heating and motion.  Due to the two types of energy transfer, these two definitions are not equivalent and muddle arguments when they are not distinguished.  The term magnetic reconnection will be used \color{black} in this paper \color{black} for changes in the magnetic field line connections.  The breaking of field line connections releases energy, which initially goes into \color{black} plasma motion involving the Alfv\'en speed, \color{black} not dissipation.  The breaking of static force balance causes the overall evolution proceed at an Alfv\'enic rate \cite{Boozer:rec2023}, \color{black} which means a rate set by the Alfv\'en speed divided by the system scale. \color{black}
 
(4)  Appendix \ref{sec:Virial} derives the Chandrasekhar and Fermi \cite{Virial:1953} virial theorem, which shows autonomous magnetic structures disassemble on an Alfv\'enic timescale unless their forces are balanced by gravity or rigid coils.

(5) Section \ref{sec:diiff} gives the difference between the magnetic field line velocity $\vec{u}_\bot$ and the plasma velocity $\vec{v}$.  Flux freezing, which is the tying together of the magnetic field lines and the plasma, has been known at least since Boozer's 2004 \emph{Reviews of Modern Physics} article  \cite{Boozer:RMP} to be two distinct conservation laws.  The better conserved of the two is the conservation of magnetic topology, such as the tying together of the toroidal and poloidal magnetic flux in toroidal plasmas.  The other is \color{black} the tying together of the  flow of the plasma and the flow of the magnetic field lines.  For example, when \color{black} $\tensor{\eta}\cdot\vec{j}=\eta_{||}\vec{j}_{||} +\eta_\bot \vec{j}_{\bot}$ with $\eta_{||}=0$ the magnetic field evolves ideally.  When $\eta_\bot$ is large, the plasma diffuses rapidly across the magnetic field lines.


\subsection{Mean-field theory}

Plasmas are almost universally in a turbulent state with two types of plasma turbulence: (i) Micro-turbulence is often on the spatial scale of the charged particle gyroradii.  (ii) Macro-turbulence is on a much larger scale, often related to the turbulence of ordinary flowing fluids.  Micro-turbulence directly modifies the entropy-producing plasma transport processes but does not fundamentally change the theory of reconnecting plasmas.  Macro-turbulence has subtle effects on magnetic reconnection. 

When the magnetic field is embedded in a macro-turbulent plasma, the magnetic field has fluctuations in both space and time on short scales and with a magnitude determined by the turbulence.  It is natural to average over the turbulence to obtain a mean-field magnetohydrodynamic theory for studying large-scale effects in plasmas.  However, a valid mean-field theory must preserve the exact conservation laws and in particular the conservation of helicity. 

The prototypical theory of mean-field magnetohydrodynamics was developed by Krause and R\"adler and summarized in their 1980 book \emph{Mean-field magnetohydordynamics and dynamo theory} \cite{Krause-Radler}.  Their best known result was that small scale turbulence modifies the large scale electric field $\vec{E}$ by adding a term $\vec{E}_T=-\alpha \vec{B} + \beta\vec{\nabla}\times\vec{B}$, much as unresolved plasma collisions add a term $\vec{E}_\eta=\eta \vec{j}$.  The coefficients $\alpha$ and $\beta$, are averages over non-linear properties of the turbulence.  The coefficient $\alpha$ became famous for it gave an explanation \color{black} for \color{black} the magnetic field in stars. 

The subtlety of mean-field theories became clear in 1986 when Boozer \cite{Boozer-mean-field:1986} noted that  the $\alpha$ effect  is inconsistent with the helicity conservation properties of low resistivity plasmas.  He also gave a form for the contribution of the turbulence to the large-scale electric field
\begin{equation} 
\vec{E}_T =  - \frac{\vec{B}}{B^2}\vec{\nabla}\cdot\left( \lambda_h \vec{\nabla} \frac{j_{||}}{B}\right). \label{Turb-E}
\end{equation}   
This form allows an arbitrarily large enhancement of the rate of reconnection and electromagnetic energy dissipation by the turbulence while conserving helicity, Section \ref{sec:helicity-ev}.  

In 1995  Bhattacharjee and Yuan \cite{Bhattacharjee-alpha:1995} derived the contribution of the turbulence to the electric field, which had the form of Equation (\ref{Turb-E}) and showed that $\alpha$ effect vanishes when the magnitude of the convective term, $\vec{v}\times\vec{B}$ is large compared to the resistive term in the standard Ohms Law. 

Helicity can be transferred from small scale turbulent eddies to the large scale field by an $\alpha$-like effect but as shown in 1999 by Ji \cite{Ji-dynamo:1999}: ``In all cases, however, the $\alpha$ effect strictly conserves the total helicity except for resistive effects and a small battery effect."  In other words, without a method to directly insert helicity into the eddies, an $\alpha$ effect cannot create helicity.  

Although helicity conservation precludes a simple Krause and R\"adler type dynamo, it does not preclude dynamos in general.  There are two possibilities.  

First, the helicity can be changed by resistivity, but the power required to produce a given rate of helicity change is enormous.  The implication is that the turbulence must produce a plasma stress tensor that transfers power from the plasma flow to the magnetic field.  The shorter the spatial scale of the turbulence the more power that must be taken from the flow to balance the resistive dissipation of magnetic energy for a given rate of helicity production.  \color{black} The convective zone in the sun produces large flows, and it  \color{black} is unclear what limit the resistive damping of the convective flow places on helicity production.    

 \color{black} Unless the plasma is evolving on an Alfv\'enic time scale, the virial theorem implies that forces associated with the magnetic field must be transferred to places where gravitational forces or rigid coils can balance them.  The turbulence-produced plasma stress tensor must be consistent with this constraint. \color{black}

Second, in 1993, Boozer \cite{Boozer-dynamo:1993} showed that a dynamo producing a large scale magnetic field must also change the magnetic helicity in regions of comparable scale.  But he also showed \cite{Boozer-dynamo:1993} that this could be accomplished by helicity transport by turbulence.  \color{black} Turbulence \color{black} can either produce separated regions of positive and negative helicity or move helicity in from the boundary.  The last option is precluded when the region is surrounded by a perfectly conducting stationary boundary.  As shown by Boozer and Elder \cite{Boozer-Elder} in 2021,  helicity can be inserted by a perfectly conducting but flowing boundary when that flow has a twist.    \color{black}

A major 2019 review of dynamo theory by Francois Rincon \cite{Dynamos:2019} pointed out the continuing use of the $\alpha$ effect in dynamo simulations.   \color{black} Rincon \color{black} appreciates the physics issues connected with the use of the $\alpha$ effect, but these issues were not emphasized in the review.  A 2023 review of the history of solar dynamos by Charbonneau and Sokoloff \cite{solar dynamo: 2023} thought the $\alpha$ effect could be restored by  helicity dissipation at small spatial scales.  However, \color{black} small scales imply large current densities, which strongly enhance energy in comparison to helicity dissipation. \color{black} Mathieu Ossendrijver \cite{Dyn-rev:2003} in a 2003 review of the solar dynamo discussed the constraint of helicity conservation in Section 3.3.2 and started Section 3.4.2 with a discussion of Interpretation of averages over fluctuations: ``Mean-field electrodynamics is a statistical theory and therefore a correct interpretation of its results requires a careful examination of the averaging procedure that is adopted." 

As the $\alpha$ effect illustrates, a critical test for an averaging procedure is whether it is consistent with invariants obtained from Maxwell's equations; helicity appears to be of particular importance.  It is unfortunate that this test is not universal.  Eyink's 2011 paper in which turbulence led to a dynamo \cite{Eyink-dynamo:2011} omits this test.  \color{black} On the other hand,  in 2014 Vishniac and Shapovalov \cite{Helicty-Vishniac:2014} emphasized the importance of helicity conservation to dynamo theory and calculated the transport of helicity by turbulence.

The review by Lazarian et al \cite{Turb-recon:2020} had seven authors and gave the collective view of an important school of research on the effect of turbulence on magnetic reconnection. \color{black} Their review has a number of insights. \color{black} Page 2 notes that ``magnetic reconnection is a ubiquitous process," which is inconsistent with requiring special conditions for reconnection to occur.  ``The problem of magnetic reconnection is not limited to explaining its typically fast rates,"  but what triggers reconnection must also be explained.  The example they give is ``flux gets accumulated prior to a solar flare and gets annihilated during the flare."  It is interesting that this particular example is easily explained by helicity conservation.  Helicity is put into a coronal loop by footpoint twisting \cite{Boozer-Elder} but has essentially zero dissipation.  When the magnetic energy in the loop that is required to be consistent with its helicity becomes too great for force balance, the loop will of necessity be ejected.  

Lazarian et al \cite{Turb-recon:2020} also note that fast reconnection is an issue of ``scale disparity.  Reconnection occurs on very large scales, while the dissipation processes take place at the smallest plasma scales, which are set by, e.g., resistivity."   Though true, the authors \color{black} discuss only turbulence produced chaos and \color{black} ignore the effect of magnetic field line chaos that arises even in smooth magnetic fields.  

On page 16, Lazarian et al \cite{Turb-recon:2020} gave a stress tensor due to turbulence that acts on the plasma.  As has been discussed, mean field theories must describe not only the effects of turbulence on the evolution of the magnetic field but also the stress tensor due to the turbulence that acts on the plasma. 

 On page 49, Lazarian et al \cite{Turb-recon:2020} make the important point: ``Most of the reconnection modeling is currently done using 2D PIC simulations. This is usually justified by the higher resolution that is available for such simulations compared to their 3D counterparts. We feel that these simulations are missing the physics essential in the large scale astrophysical reconnection processes."

Mean-field theory is clearly subtle and requires careful checking.  For example, Brandenburg and Banerjee \cite{Brandenburg:2025} in their 2025 paper discuss the effect of not only the helicity but also the Hosking integral \cite{Hosking:2021} on the decay of turbulence, which may influence the validity of a mean-field theory.  

A complete theory of magnetic reconnection in the presence of macro-turbulence is beyond the scope of this paper.  The primary effect of the turbulence on reconnection itself may well be the through the coefficient $\lambda_h$ of Equation (\ref{Turb-E}) as found in simulations of tokamak disruptions by Nardon et al \cite{Nardon:2023}. 


\subsection{Contents of the paper}

Section \ref{sec:fun-ev} is on the fundamental evolution equations.  The first three results given in Section \ref{sec:Maxwell results} are derived in this section.  

The theory of magnetic reconnection was originally developed using two-dimensional models.    Section \ref{2D-recon} explains how this resulted in three fundamental omissions: (i) magnetic field line chaos, (ii) helicity conservation, and (iii) the non-dissipative transfer of the energy out of the large scale magnetic field into \color{black} into Alfv\'enic motions, such as \color{black} Alfv\'en waves.  

The universe has three spatial dimensions.  In essentially all physical examples of reconnection, the magnetic field has a non-trivial dependence on all three, which places severe limits \color{black} on \color{black} the relevance of insights from two-dimensional reconnection models.

The most fundamental omission in two-dimensional theory was chaos.  In mathematics, a velocity $\vec{v}(\vec{x},t)$ is chaotic in a region when each streamline in that region has infinitesimally separated streamlines that increase their separation exponentially as time advances, as $\exp(t/\tau_{ef})$. Magnetic field lines are defined at an instant in time.  A magnetic field is chaotic in a region when each field line in that region has infinitesimally separated field lines that increase their separation exponentially with distance $\ell$ along the lines, as $\exp(\ell/\ell_{ef})$.   The dictionary definition of chaos, complete disorder and confusion, is not consistent with the subtle structures that can arise in chaotic systems using the mathematical definition. 

Section \ref{3D reconnection} is on reconnection in three dimensions and considers the case in which the field-line flow is on two distinct spatial scales, an eddy scale and a large scale. 

Section \ref{sec:chaos} is on magnetic-field-line chaos.  Section \ref{sec:diiff} is on the difference between the magnetic field line velocity $\vec{u}_\bot$ and the plasma velocity $\vec{v}$.  Section \ref{sec:opp-lines} explains why the Sweet-Parker model of reconnection, which is based on oppositely directed magnetic field lines, is a highly unlikely way to initiate reconnection in three dimensions.  Section \ref{sec:bndy} discusses the importance of boundary conditions on $\vec{B}$.  Section \ref{Large-scale flows} explains why large-scale flows may responsible for the dominant physical effects even when a system is highly turbulent.   Section \ref{sec:discussion} is a discussion of the paper.  Appendix \ref{sec:Virial} derives the virial theorem, which places fundamental constraints on magnetic fields that are in state of force balance.

\color{black}


\section{Fundamental evolution equations \label{sec:fun-ev}} 

\subsection{Magnetic field evolution \label{sec:B-ev}} 

The evolution of magnetic fields is rigorously given by Faraday's law, which can be written either as
\begin{eqnarray}
\frac{\partial \vec{B}}{\partial t} + \vec{\nabla}\times\vec{E} &=&0 \hspace{0.2in}\mbox{or   } \\
\frac{\partial \vec{B}}{\partial t} - \vec{\nabla}\times(\vec{u}_\bot \times \vec{B}) &=&\vec{\nabla}\ell\times\vec{\nabla}\mathcal{E}. \label{B-ev}
\end{eqnarray}
The second form for Faraday's Law follows from the purely mathematical Equation (\ref{E-eq}), which relates  two arbitrary vectors in three-space, $\vec{E}$ and $\vec{B}$ \cite{Boozer:RMP}:
\begin{eqnarray}
\vec{E}+\vec{u}_\bot \times \vec{B} &=& -\vec{\nabla}\Phi +\mathcal{E}\vec{\nabla}\ell \hspace{0.2in}\mbox{so   } \label{E-eq}\\
E_{||} &=&- \frac{\partial \Phi}{\partial\ell} + \mathcal{E}  \hspace{0.2in}\mbox{with   } \label{E|| eq} \\
\frac{\partial \mathcal{E}}{\partial\ell} &=&0.
\end{eqnarray}
$E_{||}$ is the component of the electric field parallel to the magnetic field, and $\ell$ is the distance along a magnetic field line.  $\vec{B}\cdot\vec{\nabla}\Phi= B \partial\Phi/\partial\ell$ when $(\alpha_c,\beta_c,\ell)$ Clebsch coordinates are used in which  
\begin{equation}
\vec{B}=\vec{\nabla}\alpha_c\times\vec{\nabla}\beta_c.
\end{equation}

The proof Equation (\ref{E-eq}) for $\vec{E}$ is that each of its three components can be fit anywhere that $\vec{B}$ is non-zero.  All  nulls \color{black} but point-nulls of $\vec{B}$ can be removed by an infinitesimal perturbation.  Point nulls can be addressed by placing a sphere around each null and choosing $\Phi$ on the sphere so no charge accumulates, $\oint \vec{j}\cdot d\vec{a}=0$.  \color{black} The current density $\vec{j}_0$ at a null, which is given in Equation (\ref{near-null}), automatically satisfies this condition.   \color{black}

Equation (\ref{B-ev}) not only determines the evolution for any magnetic field, but its terms also have simple interpretations.  $\mathcal{E}$ gives the departure from an ideal magnetic evolution.  When $\mathcal{E}=0$, the vector $\vec{u}_\bot$ is the velocity of magnetic field lines through space as was shown by Newcomb \cite{Newcomb} in 1958.  The proof is simple. When $\mathcal{E}=0$, Equation (\ref{B-ev}) is solved in Clebsch coordinates, in which $\vec{B}=\vec{\nabla}\alpha_c\times\vec{\nabla}\beta_c$, by 
\begin{eqnarray}
&& \frac{\partial\alpha_c}{\partial t} +\vec{u}_\bot \cdot\vec{\nabla}\alpha_c =0 \mbox{   and  } \label{d alpha/dt}\\
&& \frac{\partial\beta_c}{\partial t} +\vec{u}_\bot \cdot\vec{\nabla}\beta_c =0 \label{d beta/dt}.
\end{eqnarray}  
The labels of a magnetic field line, $\alpha_c(\vec{x},t)$ and $\beta_c(\vec{x},t)$, are carried by the flow, which implies magnetic field lines are carried by the flow and cannot break.  The magnetic flux in a region defined in Clebsch coordinates is $\int \vec{B}\cdot\vec{\nabla}\ell \mathcal{J}_cd\alpha_c d\beta_c=\int\alpha_c d\beta_c$ since the Jacobian of Clebsch coordinates $\mathcal{J}_c\equiv1/(\vec{\nabla}\alpha_c\times\vec{\nabla}\beta_c)\cdot\vec{\nabla}\ell$.  When $\mathcal{E}=0$, the magnetic flux in a tube defined by magnetic field lines cannot change.

$\mathcal{E}(\alpha_c,\beta_c,t)$ can be chosen to be zero unless the electric potential $\Phi$ must satisfy boundary conditions at two values of $\ell$ or a periodicity constraint.  An appropriate choice of $\Phi$ allows $\mathcal{E}$ to be made independent of $\ell$.  In a toroidal plasma $\mathcal{E}\vec{\nabla}\ell$ can be replaced by $(V_\ell/2\pi)\vec{\nabla}\varphi$, where $V_\ell$ is the toroidal loop voltage and $\varphi$ is any toroidal angle.  For $E_{||}=\eta j_{||}$,  $\mathcal{E}$ is proportional to $\eta$.  

When the system has open magnetic field lines, the meaning of reconnection is subtle. In one resolution, the system is surrounded by a perfect conductor through which magnetic field lines can penetrate but on which \color{black} the normal field $\vec{B}\cdot\hat{n}$ is independent of time in the frame of the conductor.  \color{black}  This resolution allows currents to close by flowing through the surrounding conductor.  \color{black} When the perfect conductor is moving at a prescibed velocity $\vec{v}_b$, the normal field on the boundary obeys $\partial(\vec{B}\cdot\hat{n})/\partial t+\vec{v}_b\cdot\vec{\nabla}(\vec{B}\cdot\hat{n})=0$. \color{black}   Another resolution is to surround the system by a perfect insulator, which means $\vec{j}\cdot\hat{n}=0$  \color{black} with the normal field to the insulator set by the solution to Laplace's equation that ensures $\vec{\nabla}\times\vec{B}=0$ and $\vec{\nabla}\cdot\vec{B}=0$ outside the insulator.\color{black}

The simple Ohm's law,
\begin{equation}
\vec{E} + \vec{v}\times\vec{B} =\eta \vec{j} \label{s-Ohms}
\end{equation}
with $\eta$ a spatial constant, and Ampere's law $\vec{\nabla}\times\vec{B} = \mu_0\vec{j}$ provide an easily interpretable approximation to the exact form of Faraday's Law, Equation (\ref{B-ev}):
\begin{equation} 
\frac{\partial \vec{B}}{\partial t} - \vec{\nabla}\times \left(\vec{v}(\vec{x},t)\times \vec{B} \right)=\frac{\eta}{\mu_0} \nabla^2\vec{B}, \label{s-B-ev}
\end{equation}   
which has the form of an advection-diffusion equation. This equation was well known when Elsasser wrote his 1956 Reviews of Modern Physics article \cite{Elsasser:1956} on hydrodynamic dynamo theory.  But, that was long before Hassan Aref \cite{Aref,Aref:PF} recognized the fundamental importance of chaos to equations of the advection diffusion form \color{black} in 1984. \color{black} An approximate equation for the magnetic evolution is Equation (\ref{s-B-ev}) but with $\vec{v}(\vec{x},t)\times \vec{B}$ interpreted as $\vec{u}_\bot\times\vec{B}$.  The $\vec{u}(\vec{x},t)$ used in the model need not have $\vec{B}\cdot\vec{u}=0$ because the cross product $\vec{u}_\bot\times\vec{B}$ removes the effect of a component parallel to $\vec{B}$.

Magnetic reconnection is commonly observed to have a timescale that is much closer to that given by the evolutionary flow $\vec{u}_\bot$ than to that defined by the $\mathcal{E}$, which is \color{black} commonly taken to be \color{black} given by resistive dissipation.  This implies that the flow is more important to determining where and when reconnection occurs than is the resistivity despite the resistivity or other non-ideal effect being required for reconnection.  As discussed in Section \ref{2D-recon}, an exception is reconnection in a two-dimensional space in which a chaotic flow produces an exponential increase in the magnetic field strength and not field line chaos.

\subsection{Magnetic helicity  evolution \label{sec:helicity-ev}} 

The conservation properties of magnetic helicity, $K$, are given by
\begin{eqnarray}
K&\equiv&\int \vec{A}\cdot\vec{B} d^3x \hspace{0.2in}\mbox{with}\hspace{0.2in}\\
\frac{dK}{dt}&=&- 2\int \vec{E}\cdot\vec{B} d^3x,\\
&=&- 2\int \mathcal{E} \vec{B}\cdot\vec{\nabla}\ell d^3x\\
&=& - 2 \int \mathcal{E} d\alpha_c d\beta_c d\ell.
\end{eqnarray}
when surface terms are ignored.  The term $\mathcal{E} \vec{B}\cdot\vec{\nabla}\ell$ can be approximated as $\eta \vec{j}\cdot\vec{B}$. The importance of helicity conservation \cite{Berger:1984} has been appreciated since Taylor's 1974 paper \cite{Taylor:1974} showing that helicity explained how strong turbulence in reversed field pinches led to quiescent periods.   The volume integral of $j_{||}B$ is insensitive the current concentrating in thin sheets. 

Helicity conservation places a strong constraint on the energy that can be released in a magnetic reconnection---even in highly turbulent plasmas.  As   proven \cite{Woltjer:1958} by L. Woltjer, \color{black} the minimum magnetic energy with fixed helicity is given by 
\begin{eqnarray}
&&\delta\left(\frac{1}{2\mu_0}\int (B^2 -\lambda_L \vec{A}\cdot\vec{B})d^3x\right) =0\\
&&\frac{1}{\mu_0}\int(\vec{\nabla}\times\vec{B}  - \lambda_L \vec{B})\delta\vec{A} d^3x=0, \hspace{0.2in}\mbox{so   }\\
&&\vec{j}=  \frac{\lambda_L}{\mu_0} \vec{B},
\end{eqnarray}
where $\lambda_L$ is a \color{black} constant, a \color{black} Lagrange multiplier.  The minimum the energy state with fixed helicity has $\vec{j}=(j_{||}/B)\vec{B}$ with the magnitude of the helicity determining \color{black} the Lagrange multiplier and \color{black} $j_{||}/B$.   

Tokamak disruptions can destroy all the magnetic surfaces on a timescale of milliseconds, helicity conservation explains why only a small drop in the energy in the poloidal magnetic field occurs and why the net toroidal current even increases \cite{Boozer:2019}. 

The non-axisymmetric turbulence associated with tokamak disruptions is of sufficient importance that it is frequently represented in axisymmetric simulation codes by the addition of a helicity-conserving term to Ohm's Law \cite{Boozer-mean-field:1986}:
\begin{equation}
 - \frac{\vec{B}}{B^2}\vec{\nabla}\cdot\left( \lambda_h \vec{\nabla} \frac{j_{||}}{B}\right). \label{helicity conserv}
 \end{equation}
The positive coefficient $\lambda_h$ is determined by the magnetic energy dissipated by the turbulence since the contribution of this term to $\vec{j}\cdot\vec{E}$ is
 \begin{equation}
\lambda_h \left(\vec{\nabla}\frac{j_{||}}{B} \right)^2-\vec{\nabla}\cdot\left(\lambda_h \frac{j_{||}}{B} \vec{\nabla}\frac{j_{||}}{B}\right).
 \end{equation} 
 

\subsection{Magnetic energy evolution \label{sec:energy-ev}}  

Maxwell's equations for $\partial \vec{B}/\partial t$  and $\partial \vec{E}/\partial t$ give Poynting's equation
\begin{eqnarray}
\frac{\partial}{\partial t} \left( \frac{B^2}{2\mu_0} + \frac{\epsilon_0 E^2}{2} \right) + \vec{\nabla}\cdot\left(\frac{\vec{E}\times\vec{B}}{\mu_0}\right) = -\vec{j}\cdot\vec{E}.
\end{eqnarray}
Ignoring surface terms, the rate of change of electromagnetic energy in a volume is
\begin{equation}
\frac{\partial}{\partial t} \int \left( \frac{B^2}{2\mu_0} + \frac{\epsilon_0 E^2}{2} \right)d^3x = - \int \vec{j}\cdot\vec{E} d^3x.
\end{equation}
The velocity of a plasma due to the electric field is $\vec{v}=(\vec{E}\times\vec{B})/B^2$, so characteristically $\mu_0\epsilon_0E^2/B^2\sim (v/c)^2$, where $\mu_0\epsilon_0=1/c^2$ with $c$ the speed of light.  In non-relativistic plasmas, the electromagnetic energy is essentially the magnetic energy, which for simplicity will be assumed.  Then,
\color{black}
\begin{eqnarray}
\frac{\partial}{\partial t} \int\frac{B^2}{2\mu_0}d^3x&=&-\int\vec{j}\cdot\vec{E} d^3x\\
&=&- \int \Big(\vec{u}_\bot\cdot\vec{f}_L + \mathcal{E} \vec{j}\cdot\vec{\nabla}\ell ) d^3x; \hspace{0.2in} \label{B-energy} \\
\vec{f}_L &\equiv& \vec{j}\times\vec{B}
\end{eqnarray}
is the Lorentz force.  The term $\vec{u}_\bot\cdot\vec{f_L}$ is the non-dissipative power and $\mathcal{E} \vec{j}\cdot\vec{\nabla}\ell$, which can be approximated by $\eta j_{||}^2$, is the dissipative power transferred out of the magnetic field.   A concentration of the current in thin sheets produces an arbitrarily large enhancement of the dissipative power transfer but not the helicity dissipation.


\section{Reconnection in two dimensions \label{2D-recon}}

The theory of magnetic reconnection developed out of two-dimensional models, 
\begin{eqnarray}
\vec{B}(x,y,t) &=& \vec{\nabla}\times \big(A_z(x,y,t)\hat{z}\big) \hspace{0.2in}\mbox{with   } \label{2D-B}\\
\vec{u}(x,y,t)&=&u_x\hat{x}+u_y\hat{y}.  
\end{eqnarray}
Equation (\ref{s-B-ev}) is then equivalent to  
\begin{equation}\frac{\partial A_z}{\partial t }+ \vec{u}\cdot\vec{\nabla}A_z = \frac{\eta}{\mu_0}\nabla^2 A_z, \label{A_z ev}
\end{equation}
which is of the standard two-dimensional advection-diffusion form that was studied by Hassan Aref  \cite{Aref,Aref:PF}.  Note, $\vec{u}\cdot\vec{\nabla}A_z=\vec{u}_\bot\cdot\vec{\nabla}A_z$ since $\vec{B}\cdot\vec{\nabla}A_z=0$.  

A standard example of an advection-diffusion problem has $A_z$ replaced by $T$, the temperature in a room, and $\eta/\mu_0$ replaced by $D_T$, \color{black} the thermal-diffusion coefficient in air.  This equation explains \cite{Boozer:entropy2021} why the temperature in a room becomes uniform in tens of minutes rather than in the few weeks as expected from thermal diffusion alone.

Aref's 1984 paper \cite{Aref} was the first to recognize the importance of chaos in the theory of mixing.  Although magnetic field lines cannot be chaotic in two dimensions, velocities generally are in natural flows.  Three coordinates are required for chaos, which can be $(x,y,t)$ for the velocity,  but magnetic field lines are defined at  instants \color{black} in time, so three spatial coordinates are required.

The universe has three spatial dimensions and two-dimensional models of reconnection have led to many misconceptions:

(1)  The exclusion of chaos from reconnection theory.  Magnetic field lines cannot be chaotic in two-dimensions.  Although their evolution velocity $\vec{u}$ can be, a chaotic ideal evolution would lead to an exponential increase in the field strength, which would require an exponentially large force driving the reconnection.

(2) The exclusion of magnetic helicity from reconnection theory.  The helicity is identically zero when $\vec{B} = \vec{\nabla}\times \big(A_z\hat{z}\big)$ and therefore irrelevant.  

The addition of a guide field, a magnetic field in the $\hat{z}$ direction,  gives a non-zero helicity \color{black} even when there is a dependence on only two spatial coordinates.  The standard model is the in-plane reconnecting field changes sign across a narrow region $\Delta$, which can be represented by a vector potential 
\begin{equation}
A_z = \frac{B_r}{2} x \frac{e^{x/\Delta} - e^{-x/\Delta}}{e^{x/\Delta} + e^{-x/\Delta}},
\end{equation}
with $B_r$ a constant, the magnitude of the positive and the negative reconnecting magnetic fields, and the helicity given by $\int B_z A_z dx dy dz$.  The total helicity is proportional to the width of the region in which the reconnection is occurring.  \color{black} The helicity \color{black} can be conserved by firing helicity-containing plasmoids out along the narrow region in which magnetic field changes signs.

\color{black} The conservation of helicity during reconnections cannot always be achieved by plasmoid ejection. \color{black}   As shown by Boozer and Elder \cite{Boozer-Elder}, a  twist in the footpoint motion injects helicity into individual tubes of magnetic field lines.  Fast magnetic reconnections can spread the helicity but not dissipate it \cite{Boozer-Elder}. Tubes of magnetic field lines cannot remain in force balance with an arbitrarily large helicity. \color{black} When force balance is lost in a magnetic flux tube, it is ejected. \color{black}

(3) Energy transfer in magnetic reconnection.  In two-dimensional theory, reconnection is often defined as a dissipative transfer of magnetic energy into plasma thermal energy---not a topology change, the breaking of magnetic field line connections.  In two dimensions, the dissipative transfer of energy appears to be equivalent but simpler.   As discussed in Section \ref{sec:B-ev}, a definition based on a topology change generally requires boundary or periodicity conditions to make $\mathcal{E}\neq0$.   In three dimensions, the change in topology causes little energy dissipation when $\eta$ is small, but the non-dissipative energy transfer $\vec{u}_\bot\cdot\vec{f}_L$ drives \color{black} Alfv\'enic plasma motions, such as \color{black} Alfv\'en waves, which are quickly dissipated in a chaotic magnetic field \cite{Heyvaerts-Priest:1983,Similon:1989,Boozer-Alfven:2005}.   Lazarian \cite{Lazarian} has studied the damping of Alfv\'en waves when the plasma is turbulent.  The damping depends on the wavelength of the Alfv\'en wave compared to the spatial scale of the turbulence.

When the resistivity $\eta$ is zero, Equation (\ref{A_z ev}) for $A_z$ is easily solved since $A_z$ has a fixed value in a frame carried by the flow.  This is the frame of $(x_0,y_0)$  of \color{black} Lagrangian coordinates; $dA_z/dt \equiv (\partial A_z/\partial t) + (d\vec{x}/dt)\cdot\vec{\nabla}A_z =0$ where $d\vec{x}/dt=\vec{u}$.  

In two dimensions, Lagrangian coordinates are defined so ordinary Cartesian coordinates $(x,y)$ obey $\partial x(x_0,y_0,t)/\partial t = u_x(x,y,t)$ and $\partial y(x_0,y_0,t)/\partial t = u_y(x,y,t)$ with a $t=0$ initial condition that $x=x_0$ and $y=y_0$.  In Lagrangian coordinates, $A_z(x_0,y_0)$ is independent of time, so
\begin{eqnarray}
\vec{\nabla}A_z &=& \left(\frac{\partial A_z}{\partial x_0} \frac{\partial x_0}{\partial x} + \frac{\partial A_z}{\partial y_0} \frac{\partial y_0}{\partial x}\right)\hat{x} \nonumber\\  && +  \left(\frac{\partial A_z}{\partial x_0} \frac{\partial x_0}{\partial y} + \frac{\partial A_z}{\partial y_0} \frac{\partial y_0}{\partial y}\right)\hat{y} \hspace{0.2in}\mbox{so} \hspace{0.2in} \\ \nonumber\\
 \left(
\begin{array}{ccc}
  \frac{\partial A_z}{\partial x}     \\
 \frac{\partial A_z}{\partial y} 
\end{array}
\right) &=& \left(
\begin{array}{ccc}
  \frac{\partial x_0}{\partial x}&  \frac{\partial y_0}{\partial x}     \\
 \frac{\partial x_0}{\partial y} &\frac{\partial y_0}{\partial y} 
\end{array}
\right) \cdot\left(
\begin{array}{ccc}
  \frac{\partial A_z}{\partial x_0}     \\
 \frac{\partial A_z}{\partial y_0} 
\end{array}
\right) \\ \nonumber\\
B^2 &=&(\vec{\nabla}A_z)\cdot (\vec{\nabla}A_z)\\
&=& \left(
\begin{array}{ccc}
  \frac{\partial A_z}{\partial x}     \\
 \frac{\partial A_z}{\partial y} 
\end{array}
\right)^\dag \cdot \left(
\begin{array}{ccc}
  \frac{\partial A_z}{\partial x}     \\
 \frac{\partial A_z}{\partial y} 
\end{array}
\right)
\end{eqnarray} 
The matrix
\begin{eqnarray}
\tensor{J}_L\equiv \left(
\begin{array}{ccc}
  \frac{\partial x_0}{\partial x}&  \frac{\partial y_0}{\partial x}     \\
 \frac{\partial x_0}{\partial y} &\frac{\partial y_0}{\partial y} 
\end{array}
\right),
\end{eqnarray}
is called the Jacobian matrix of Lagrangian coordinates.   In a singular value decomposition, $\tensor{J}_L$ has two singular values, which must have a product of unity when the flow is divergence free.  When a divergence-free flow is chaotic, one singular value has a positive exponential dependence on time, as $\exp(t/\tau_{ef})$, and the other a negative, $\exp(-t/\tau_{ef})$, where $\tau_{ef}$ is the timescale for an e-fold in separation between neighboring streamlines.

\color{black}  In the two dimensional case the current density is precisely perpendicular to the field, and the velocity $\vec{u}_\bot$ is perfectly aligned with the Lorentz force. Both effects maximize the $\vec{u}_\bot\cdot\vec{f}_L$ power transfer.  The Lorentz force $\vec{f}_L\equiv\vec{j}\times\vec{B}$ and the perpendicular velocity are given by
\begin{eqnarray} 
\vec{f}_L &=& j_z \vec{\nabla}A_z \hspace{0.2in}\mbox{where   }\\
 \vec{j} &=&j_z \hat{z} =- \hat{z} \nabla^2A_z/\mu_0 \hspace{0.2in}\mbox{and   }\\
 \vec{u}_\bot &=& \frac{\vec{\nabla}A_z}{|\vec{\nabla}A_z|^2} \vec{u}\cdot\vec{\nabla}A_z. 
 \end{eqnarray}
 \color{black}

In two dimensions, an ideal magnetic evolution with a chaotic evolution velocity leads to an exponentially large magnetic field strength unless the field is perfectly aligned with the direction of exponential decrease.  A chaotic velocity is not an explanation for fast reconnection in two-dimensions.   Indeed, two-dimensional reconnection studies are based on a non-chaotic evolution velocity, typically with two regions of constant but oppositely directed magnetic field that are pushed together.  The lack of relevance of this Sweet-Parker solution is the topic of Section \ref{sec:opp-lines}.


\section{Reconnection in three dimensions \label{3D reconnection} } \color{black}

The properties of the magnetic evolution equation, change fundamentally when the evolution velocity $\vec{u}_\bot$ depends on the third coordinate, $z$.   The exponential increase in the field strength, which is characteristic of two-dimensional chaotic flows, is not required in three dimensions. In three dimensional space, the Jacobian tensor of Lagrangian coordinates is a three-by-three tensor, which has three singular values.  For a divergence-free flow the product of these singular values must be unity.  For a chaotic flow, the largest singular value increases exponentially in time, the smallest decreases exponentially, but the third only has an algebraic dependence on time \cite{Boozer:entropy2021}.

Chaotic flows can be smooth with essentially only one spatial scale defining its variation across the magnetic field lines.   An example of a non-turbulent but chaotic flow is given by $\vec{u} = \hat{z}\times \vec{\nabla} \big((z/a_z) h\big)$, where $a_z$ is the height of a box like region that is of length $a$ and with $h$ given by Equation (\ref{stream function}). 

Here the effect of having two spatial scales in the flow will be studied: a short scale $a$ and a long scale $L$.  A far more realistic study could be carried out using a chaotic velocity $\vec{u}_\bot$ in the approximate model of magnetic field evolution, which is described in the discussion of Equation (\ref{s-B-ev}).  There is a subtlety. The presence of a magnetic field affects the flow.  The assumed flow $\vec{u}_\bot(\vec{x},t)$ must be adjusted so that at least in some average sense
\begin{equation}
|\vec{u}_\bot\cdot \vec{f}_L| << \frac{ \frac{B^2}{2\mu_0} }{\tau_{ef}}  \label{f_L inequality}
\end{equation} to avoid excessive energy input into the magnetic field before significant reconnection has occurred.   If $\vec{u}_\bot$ were perfectly aligned with $\vec{j}_\bot$, the required power $|\vec{u}_\bot\cdot \vec{f}_L|$ would be zero.   Flows naturally proceed in the direction of least back reaction.  A direction in which a minimal back force arises is possible in three dimensions but not two. \color{black}

Although the third singular value in a singular value decomposition of the Jacobian matrix does not depend exponentially on time, it does have time dependence.  When a magnetic field is forced to evolve, some power is generally required.  The required power is estimated in Section \ref{sec:power-req}.

 In three dimensions, the current density can be almost parallel to the magnetic field, which makes the Lorentz force small.  This is the case throughout a volume in which the magnetic evolution is due to a force at the boundary, the field lines are moved slowly compared to the Alfv\'en speed, and the plasma pressure is zero.  In addition, there are two directions in which $\vec{u}_\bot$ can lie and produce the advective term $\vec{u}_\bot\times\vec{B}$, which \color{black} unlike the two-dimensional case \color{black} implies $\vec{u}_\bot$ need not be perfectly aligned with the Lorentz force.

\subsection{Effect of a field-line flow on $\vec{B}$ evolution \label{sec:flow effects}} 

The flow of magnetic field lines $\vec{u}_\bot(\vec{x},t)$ may be too complicated to calculate in a highly turbulent plasma.   In some cases of interest, the finiteness of the mean free path even makes the plasma velocity $\vec{v}_\bot(\vec{x},t)$ ill defined, but it is not the actual plasma velocity but the velocity of the magnetic field lines $\vec{u}_\bot$ that determines the magnetic evolution.  Their difference is the topic of Section \ref{sec:diiff}.  At each point in space and time, a magnetic field line velocity exists, which would give the evolution of the magnetic field if $\mathcal{E}$, which is determined by non-ideal effects, such as the resistivity $\eta$, were zero.  


\begin{figure}
\centerline{ \includegraphics[width=3.2 in]{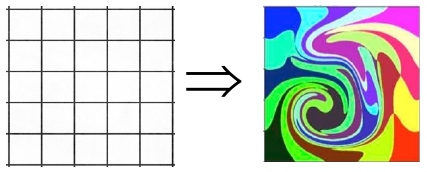}}
\caption{ \color{black}Any magnetic field $\vec{B}(\vec{x},t)$ can be thought of as consisting of tubes of magnetic flux by placing a gridded surface across the field.  Each tube is defined by the magnetic field lines that pass through the perimeters of the grid cells.  When the field is chaotic, the perimeter of each cell becomes exponentially longer when the grid is replotted after each line on the perimeters is followed for a distance $\ell$.   But, each cell contains exactly the same field lines and has precisely the same neighboring cells.  When the magnetic field is evolving ideally with a chaotic velocity $\vec{u}_\bot$, a similar distortion of the grid occurs when the grid is replotted using the location of each line on the perimeters after a time $t$.  The figure shows the distortion of a $5\times5$ array.   The distorted grid is part of Figure 5 of Y.-M. Huang and A. Bhattacharjee, Phys. Plasmas 29, 122902 (2022), which was based on a chaotic evolution defined by A. H. Boozer and T. Elder, Phys. Plasmas \textbf{28}, 062303 (2021).  Boozer and Elder illustrated distortions of flux tubes up to a factor $\sim 10^7$. \color{black}   } 
\label{fig:tubes}
\end{figure}

Using the approximate Equation (\ref{s-B-ev}) for the magnetic evolution, the term $(\eta/\mu_0) \nabla^2\vec{B}$ in Equation (\ref{s-B-ev}) must  in some sense become comparable to the ideal evolution timescale of $\vec{B}$ for some points in space for reconnection.  This statement is surprisingly subtle.  A tube of magnetic field lines, which we call a flux tube, is defined by the curve it forms on any surface that cuts across the magnetic field lines, \color{black} Figure \ref{fig:tubes}. \color{black} When $\mathcal{E}=0$, the equation $\vec{\nabla}\cdot\vec{B}=0$ implies that the magnetic flux enclosed by each defining curve located anywhere and at any time along a flux tube remains the same.  When the field is and remains chaotic, field lines infinitesimally separated from a line will exponentially separate from it with distance along a line, Section \ref{sec:chaos}.  The result is that the circumference of each of these defining curves increases exponentially in time \color{black} and in distance along the lines \color{black} without limit.  That is the defining curves become exponentially more crumpled while exactly conserving its enclosed flux.  \color{black} The flux in the $i^{th}$ cell is $\psi_i=\int \vec{B}\cdot d\vec{a} =\oint \vec{A}\cdot d\vec{s}_c$ \color{black} with $\vec{A}$ the vector potential and $ d\vec{s}_c$ the differential distance along \color{black} the curve that defines the perimeter of the cell. \color{black}  

By following the path of least resistance (meaning back force not electrical resistance), the flow $\vec{u}_\bot$ will naturally accommodate so it satisfies Equation(\ref{f_L inequality}).   Electrical resistivity anywhere along a field line, $(\eta/\mu_0)\nabla^2\vec{B}$ causes diffusion, which creates uncertainty  as time advances in the position of the field line.  \color{black} Because of the exponential increase of small separations in chaotic fields,  the region over which the future position a line is uncertain due to $(\eta/\mu_0)\nabla^2\vec{B}$ at a particular point becomes greater the larger the distance from that $(\eta/\mu_0)\nabla^2\vec{B}$  point.  The places of highest crumpling are the most sensitive to this uncertainty.  Mixing field lines from within a flux tube from those without is a breaking of topology and hence a magnetic reconnection everywhere along the lines that are intermixed.  The reconnection in some sense takes place at the point along the boundary of the flux tube that becomes most crumpled even if far from the place along the line where $(\eta/\mu_0)\nabla^2\vec{B}$ is maximized.

\color{black} 

Let $a_\bot$ be the initial characteristic spatial scale across the magnetic field lines.  Field lines that become closer than $\delta$ will reconnect when the rate of field line diffusion $|(\eta/\mu_0) \nabla^2\vec{B}| \approx (\eta/\mu_0)  B/\delta^2$ is comparable to one over the minimum timescale $(a_\bot/\delta)\tau_{ef}$ required for two lines initially separated by $a_\bot$ to have a separation $\delta$. As discussed, reconnection on the scale $a_\bot$ occurs in very specific places.  The resistive field-line interfusion only occurs where two field lines that are initially separated by $a_\bot$ are carried by the flow to a separation $\delta$ with $\delta/a_\bot \approx 1/\mathcal{R}_a$,  where
\begin{equation}
\mathcal{R}_a \equiv \frac{(\mu_0/\eta)a_\bot^2}{\tau_{ef}}. \label{R_a} 
\end{equation}  \color{black}   
$\mathcal{R}_a$ is the resistive timescale $\tau_\eta=a_\bot^2/(\eta/\mu_0)$ divided by $\tau_{ef}$  and is often comparable to what many authors call the magnetic Reynold number in which $\tau_{ef}$ is replaced by $a_\bot/v$ with $v$ the plasma flow speed.  

 As time advances, the grid cells \color{black} in  Figure \ref{fig:tubes} distort until $\mathcal{E}$, no matter how small it may be, will intermix flux from the different cells and produce a large scale magnetic reconnection.  

\color{black} 

An initially simple magnetic field must evolve for a time $\approx\tau_{ef} \ln(\mathcal{R}_a)$ before significant reconnection occurs.   In reconnection problems of practical interest, $\mathcal{R}_a$ has values from $10^4$ to $10^{20}$, and the actual value of $\mathcal{R}_a$ makes only a small change in $\ln(\mathcal{R}_a)$ over this range. The natural logarithm of $\mathcal{R}_a$ changes from 9.2 to 46, which is a total variation of a factor of five.  The factor $\ln(\mathcal{R}_a)$ differs from twenty by approximately a factor of two even as $\mathcal{R}_a$ varies by sixteen orders of magnitude. 


\color{black}  Eric Priest and collaborators \cite{Priest:2021} have stressed the importance of nulls, $\vec{B}=0$ points, in astrophysical reconnection.  Chaos as a cause for reconnection may be more intuitive in situations like tokamak disruptions where no magnetic nulls exist.  But, chaos may also be central to reconnection in the presence of nulls.  As Elder and Boozer \cite{Elder-Boozer 2021} have discussed, near a null a magnetic field has the Taylor expansion 
\begin{equation}
\vec{B}(\vec{x}) = \tensor{M}\cdot\vec{x} + (\mu_0/2)\vec{j}_0\times\vec{x}, \label{near-null}
\end{equation} 
where $\tensor{M}$ is a symmetric traceless matrix and $\vec{j}_0$ is the current density at  the null.  The implied Lorentz force at the null has a curl, $\vec{\nabla}\times\vec{f}_L = -\vec{j}_0\cdot\tensor{M}$, which is presumably balanced by the inertia associated with imparting vorticity to the plasma.  This would produce a large distortion to the tubes of magnetic flux that pass by the null.  There is a related discussion by Pontin,  Bhattacharjee, and Galsgaard \cite{Pontin:2007} in 2007.

\color{black}

\subsection{Power required to maintain the field line flow \label{sec:power-req}} 

The e-folding of magnetic field line separation and the power required to drive the field line flow velocity $\vec{u}_\bot$ are related.  The separation between two infinitesimally separated field lines depends on distance along the lines as $d\vec{\delta}/d\ell = (\vec{\delta}\cdot \vec{\nabla})\hat{b}$ where $\hat{b}\equiv \vec{B}/|\vec{B}|$.  When the field lines are chaotic, the characteristic e-folding distance is $\ell_{ef}\approx 1/|\vec{\nabla}\hat{b}|$.  The current density required to produce $|\vec{\nabla}\hat{b}|$ is primarily along the magnetic field and of order $\mu_0j_{rq}/B\approx|\vec{\nabla}\hat{b}|$.  Consequently, $j_{rq} \approx B/\mu_0\ell_{ef}$.  To obtain reconnection, $\mathcal{R}_a$ e-folds are required, where $\mathcal{R}_a$   is defined in Equation (\ref{R_a}).  This implies \color{black} the exponentiation must persist for  a longer scale along the lines than $a_{||}=\ell_{ef} \ln(\mathcal{R}_a)$. \color{black}  When the parallel current $j_{rq}$ varies on the spatial scale $a_{||}$ along $\vec{B}$,  \color{black} the divergence of the parallel current produces a perpendicular current \color{black} $j_\bot\approx (\ell_{ef}/a_{||}) j_{rq}$ \color{black} and a Lorentz force $f_L\approx (\ell_{ef}/a_{||}) j_{rq}B$.  The reference-frame independent velocity $u_\bot$ is $\ell_{ef}/\tau_{ef}$.   The required power input to the magnetic field to maintain the flow is then $P_{rq} \approx (\ell_{ef}/\tau_{ef}) f_L$.   \color{black} The implication is that the magnetic field must change by order itself during the time $\tau_{rec}\approx \tau_{ef}\ln(\mathcal{R}_a)$ for reconnection to occur since 
\begin{equation} 
P_{rq} \approx \frac{B^2}{\mu_0 \tau_{rec}},  \label{rq-power}
\end{equation}
which is consistent with Equation(\ref{f_L inequality}) since it is smaller by a factor of $1/\ln(\mathcal{R}_a)$. \color{black}  The power being resistively dissipated is $P_\eta\approx\eta j_{rq}^2$.  The ratio
\begin{eqnarray}
\frac{ P_{rq}}{P_\eta} &\approx&  \frac{1}{\ln(\mathcal{R}_a)} \frac{(\mu_0/\eta)a_\bot^2}{\tau_{ef}} \\
& \approx& \frac{\mathcal{R}_a}{\ln(\mathcal{R}_a)}.
\end{eqnarray}
The required power input is enormous compared to that dissipated by resistivity when $\mathcal{R}_a\rightarrow\infty$  even though the required power input to distort the tubes of flux, \color{black} Equation(\ref{rq-power}), is relatively  \color{black} small.  As discussed in Section \ref{sec:flow effects}, the flow $\vec{u}_\bot$ naturally accommodates to minimize the power required to produce the flux tube distortions. 

\color{black}

Helicity conservation limits the energy that can be released by a reconnection.  Nonetheless, the released energy is enormous compared to that that can be resistively dissipated and must the transferred out of the large scale magnetic field by the non-dissipative term $\vec{u}_\bot\cdot\vec{f}_L$ in Equation (\ref{B-energy}) for the time derivative of the magnetic energy.  When the Lorentz force $\vec{f}_L$ is not balanced by static forces, such as $\vec{\nabla}p$ in an equilibrium, $\vec{f}_L$ must be balanced by inertial forces, which means \color{black}  plasma motion involving the Alfv\'en speed.  \color{black}  As noted,  shear \color{black} Alfv\'en waves are quickly damped by forming thin sheets of vorticity and current.

\subsection{Reconnection with small scale eddies}

The description of magnetic reconnection of tubes of magnetic flux, Section \ref{sec:power-req}, has two spatial scales.  These are $a_\bot\approx \ell_{ef}$ across and the other $a_{||}$ along  $\vec{B}$.  Their ratio is $a_{||}/a_\bot\approx \ln(\mathcal{R}_a)$.   The size of $a_{||}/a_\bot$ implies the \color{black} the exponential amplification \color{black}   mentioned in Section \ref{sec:flow effects} is large.  When the flow $\vec{u}_\bot$ has small eddies as well as a large scale flow, the theory is unchanged as long as both $a_\bot$ and $a_{||}$ are shorter than the related scales of the eddies.  An eddy will generally have different scales along and perpendicular to $\vec{B}$, and either can be the more restrictive scale on $a_{||}$ or $a_\bot$. 

\color{black}

If reconnection occurred only on the scale of the eddies, magnetic field lines would have to be followed a long distance $L_{||}$ before moving a distance comparable to the largest scale of the flow across magnetic field lines, $L_f$.  A field line must be followed a distance $a_{||}$ to move a distance $a_\bot$ across the field.  These $a_\bot$ steps are a random walk, so $(L_f/a_\bot)^2$ steps are required.  Consequently, the total distance field lines must be followed to cross the field by the distance $L_f$ is $L_{||} \approx a_{||} (L_f/a_\bot)^2$.  This distance is of order $L_f/a_\bot$ longer than if the reconnection had been on the scale $L_f$ instead of $a_\bot$.  When the large scale flow is chaotic, with an e-folding time $\tau_{EF}$, that flow will carry lines in ways that produce the larger scale reconnection on a timescale $\tau_{EF} \ln{\mathcal{R}_L}$,  where $\mathcal{R}_L\approx(\mu_0/\eta)L_f^2/\tau_{EF}$.  The distance a field line must be followed to cross the region of width $L_f$ is 
\begin{equation}
 L_{||} \approx L_f \ln\left(\frac{a_{||} \left(\frac{L_f}{a_\bot}\right)^2}{\ell_{EF}}\right),
 \end{equation}
 where $\ell_{EF}$ is the characteristic e-folding distance of field lines in the large scale flow.


\section{Chaotic magnetic field lines \label{sec:chaos}}

The effect of chaos on magnetic reconnection depends primarily on whether the magnetic field lines become chaotic as they evolve.  Magnetic field lines are defined at a particular  instant \color{black} and are chaotic in a region of space when each line has lines in its neighborhood that exponentially separate from it with distance $\ell$ along the line. \color{black}  

Magnetic field lines that are infinitesimally separated from an arbitrarily chosen line by a distance $\rho$ are given by a Hamiltonian \cite{Boozer:line-sep}
\begin{eqnarray}
H(\psi,\alpha,\ell)=\Big(k_\omega(\ell) + k_q(\ell)\cos(2\alpha - 2\varphi_q(\ell))\Big)\psi, \label{Local-H} \hspace{0.1in}
\end{eqnarray}
where $\psi\equiv B_0(\ell)\rho^2/2$, $d\psi/d\ell=-\partial H/\partial \alpha$, and $d\alpha/d\ell=\partial H/\partial \ell$.  There are four functions of distance $\ell$ along the chosen line: $B_0(\ell)$, the field strength along the line, $k_\omega(\ell) =\tau_0(\ell) +(\mu_0j_{||}/B)_0$, where $\tau_0$, the torsion of the line, and $(\mu_0j_{||}/B)_0$ are given along the line; $k_q(\ell)$ and $\varphi(\ell)$ are the strength and phase of the quadrupole term in a Taylor expansion of the magnetic field in $\rho$.  All four functions of $\ell$ can evolve in an ideal evolution.  When $k_q(\ell)$ is sufficiently large and rapidly varying compared to $k_\omega$, field lines near the chosen line have a separation that is changing exponentially.  $\vec{\nabla}\cdot\vec{B}=0$ ensures neighboring lines must have both signs of exponentiation, which means separation or approach.  The trajectories of the Hamiltonian of Equation (\ref{Local-H}) are easily shown to have this property, $d\ln(\psi)/d\ell =2 k_q(\ell)\sin\big(2\alpha - 2\varphi_q(\ell)\big).$ 


\begin{figure}
\centerline{ \includegraphics[width=3 in]{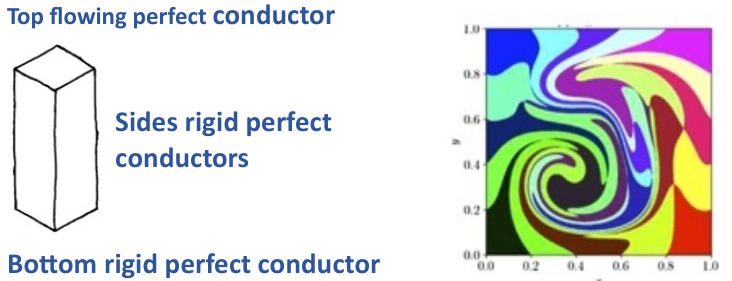}}
\caption{\color{black} This figure illustrates how the flux tube distortion illustrated in Figure \ref{fig:tubes} can arise in a model of coronal loops in which the evolution is driven by footpoint motion.  \color{black}  Magnetic flux tubes in a perfectly conducting plasma are shown that are bounded by a perfectly conducting box with rigid sides and a flowing top.  On the bottom of the box, the tubes are five by five squares.  The intersection of these tubes with the top is part of Figure 5 of Y.-M. Huang and A. Bhattacharjee, Phys. Plasmas 29, 122902 (2022).  The tubes are illustrated after a sufficiently long time to become highly distorted but before resistive diffusion has had significant effects.   } 
\label{fig:dist}
\end{figure}

Even when $\eta=0$, an evolving magnetic field in three dimensions will generally become chaotic.    Figure \ref{fig:dist} illustrates the behavior of magnetic flux tubes in a system with negligible resistivity and well-defined but evolving boundary conditions. The walls of the rectangular box are all perfectly conducting.  The side walls at $x=\pm a$ and $y=\pm a$ as well as the bottom wall at $z=0$ are stationary.  The top wall at $z=L$ is perfectly conducting but has a divergence-free flow $\vec{v}_L = \hat{z} \times \vec{\nabla} h(x,y,t)$ given by a stream function, Equation (\ref{stream function}).  The gradient of the stream function, $\vec{\nabla}h$, that was used to produce Figure \ref{fig:dist}  vanished near $x=\pm a$ and $y=\pm a$.  This ensured the velocity of the top was continuous with the zero velocity of the side walls. Otherwise $h$ is proportional to a function $\tilde{h}$ that is a Taylor series in $x$ and $y$ times sine and cosine oscillatory functions of time:
\begin{eqnarray}
&& h = \tilde{h}(x,y,t) \left(1-\frac{x^2}{a^2}\right)^3\left(1-\frac{y^2}{a^2}\right)^3; \label{stream function}\\
&&\tilde{h} =\frac{a^2}{\tau}\Big\{ c_0 \cos\left(\omega_0\frac{t}{\tau}\right) + c_1 \frac{x}{a} \cos\left(\omega_1\frac{t}{\tau}\right) \nonumber\\
&& \hspace{0.3in} +  c_2 \frac{y}{a} \cos\left(\omega_2\frac{t}{\tau}\right) + c_3 \frac{xy}{a^2} \cos\left(\omega_3\frac{t}{\tau}\right)\Big\}. \hspace{0.2in} \label{h-form}
\end{eqnarray}
In the integrations to determine the distorted flux tubes of Figure \ref{fig:dist},  $c_0=0$, which implies no magnetic helicity is injected into the rectangular box.  Magnetic helicity cannot be dissipated faster than the global resistive timescale while magnetic energy and topology can be destroyed on a timescale only $\ln(\mathcal{R}_a)$ longer that $\tau_{ef}$.  Integrations with $c_0=0$ produce only those effects that can rapidly dissipated.  The other constants were $c_1=c_2=c_3=1/2$, $\omega_1=6\pi$, $\omega_2=4\pi$, and $\omega_3=0$.  These choices give a field line separation that exponentiates with $\tau_{ef} \approx \tau$.  

The flow remains chaotic for many choices of the $c'$s and $\omega'$s, but at least one $\omega$ must be non-zero.  When the stream function $h$ is independent of time, $h$ is a constant of the motion, which is inconsistent with chaos.

Flux tubes were defined in Figure \ref{fig:dist} by dividing the $z=0$ surface into $5\times5$ squares \color{black} as they were in Figure \ref{fig:tubes}.  \color{black}  $\vec{\nabla}\cdot\vec{B}=0$ implies exactly the same flux passes through any surface that crosses the tube.  The flux tubes plotted in Figure \ref{fig:dist} are the distortions that the $5\times5$ tubes that were squares at $z=0$ have undergone by the time they reach the top $z=L$ after a few e-folds.    Note the tubes are undistorted near the side walls because the velocity of the top surface vanishes there.

The distortions to the flux tubes illustrated in Figure \ref{fig:dist} increase exponentially in time.  The closest approach of distinct flux tubes becomes exponentially smaller.  An arbitrarily small resistive diffusion $\eta/\mu_0$ can change field line connections by interdiffusing field lines from different tubes after a time that depends only logarithmically on $\eta/\mu_0$. 

Pondering the implications of  Figures \color{black} \ref{fig:tubes} and \color{black} \ref{fig:dist} leads to five important conclusions:   \color{black}

(1)   An ideal evolution generally makes magnetic field lines chaotic with the exponentiation of infinitesimally separated lines increasing on the timescale $\tau_{ef}$ of the flow.

(2) The increasing chaos ends any approximation to magnetic field lines preserving connections on a  timescale that depends only logarithmically on $\eta/\mu_0$.  This timescale  can be approximated as twenty times longer than $\tau_{ef}$ and is almost independent of the diffusion coefficient, $\eta/\mu_0$, when the timescale for diffusion is very long compared to $\tau_{ef}$.\color{black}

(3)  The average rate of exponentiation, which is called the Lyapunov exponent, can be zero with the exponentiation remaining important.  This is the case if the moving surface \color{black} in Figure \ref{fig:dist} \color{black} is placed halfway up in the tall box with all six outer surfaces of the box rigid perfect conductors.

(4) The breaking of field line connections, reconnection, occurs both due to the perimeter of the flux tubes increasing exponentially in time and the shortest distance between different tubes decreasing exponentially in time.

(5)  When reconnection occurs due to the decrease in the shortest distance between different tubes, the reconnection occurs at very specific locations---a retying of field lines after an almost scissor-like cutting.


\section{Difference between field line and plasma velocities \label{sec:diiff}}

The magnetic field line velocity $\vec{u}_\bot$ is distinct from the plasma velocity $\vec{v}$.  This difference has two parts.  The most obvious difference is that  $\vec{v}=\vec{v}_\bot + \vec{v}_{||}$ while the magnetic field line velocity is only defined perpendicular to the field lines, $\vec{u}_\bot$.  

The plasma can also have a different velocity perpendicular to the magnetic field lines $\vec{v}_\bot$ than $\vec{u}_\bot$.  The difference $\vec{v}_\bot-\vec{u}_\bot$ is given by subtracting Equation (\ref{E-eq}) for general electric field from the electric field of Ohm's law.  The conventional Ohm's law plus the Hall term due to the Lorentz force, $\vec{f}_L=\vec{j}\times\vec{B}$, exerted by the magnetic field is
\begin{eqnarray}
 \vec{E} + \vec{v}\times\vec{B} &=& \eta_{||}\vec{j}_{||}+\eta_\bot\vec{j}_{\bot} +\frac{\vec{f}_L}{n_e e}, \mbox{   so   } \label{Hall-Ohm's}\\
 E_{||}&=&\eta_{||} j_{||},
\end{eqnarray} 
where $n_e$ is the electron number density.  This equation for $E_{||}$ and Equation (\ref{E|| eq}), which relates $\mathcal{E}$ to $E_{||}$, imply the  Hall term in Ohm's law has no direct effect on $\mathcal{E}$ and, therefore, no direct effect on the breaking of magnetic field line connections.  This is despite what is claimed in many papers on reconnection.  But, the Hall term  does affect the flow of the plasma across the magnetic field lines, $\vec{v}_\bot-\vec{u}_\bot$:
\begin{eqnarray}
(\vec{v}_\bot-\vec{u}_\bot)\times\vec{B}= && \vec{\nabla}\Phi +\frac{\vec{f}_L}{n_e e} \nonumber\\
 && + \eta_{||}\vec{j}_{||}+\eta_\bot\vec{j}_{\bot} -\mathcal{E}\vec{\nabla}\ell. \hspace{0.2in} \label{Plasma-B vel}
\end{eqnarray}

The difference in the plasma and the field line flow across the magnetic field is due both to ideal terms, which are in the upper line of the right-hand side of Equation (\ref{Plasma-B vel}), and dissipative terms, which are given in the lower line.  The ideal terms have a typical magnitude  $\sim T/ea$, where $T$ is the plasma temperature $a$ is a characteristic distance.  

The electric potential $\Phi$ is approximately $T/e$ because that is what is required to maintain the quasi-neutrality of the plasma when the ions and electrons have any differences in transport magnitudes.  $\Phi$ must balance the pressure of the poorer confined species.  The Hall term is often approximately given by $\vec{\nabla}p/en_e$, which is itself approximated by $T/ea$.  The difference $\vec{v}_\bot-\vec{u}_\bot$ can be small in astrophysical plasmas though it is often large in laboratory plasmas.  \color{black}  In addition, the plasma inertial force $\rho_m \partial \vec{v}/\partial t$ makes the Hall term sufficiently large to disconnect $\vec{v}_\bot$ and $\vec{u}_\bot$ when  $\vec{v}_\bot$ changes at a rate comparable to the ion cyclotron frequency.   \color{black}  

Even when $\vec{v}_\bot$ and $\vec{u}_\bot$ are approximately equal, the evolution of the magnetic field and the plasma properties can be fundamentally different because the plasma motion includes a parallel flow and the field line motion does not.  The important quantity in determining the nature of the solutions of advection-diffusion equation with a small diffusion is not the diffusion but the $\tau_{ef}$, the time required for the streamlines of the flow to go through an e-fold of separation.



\section{Reconnection model of oppositely directed lines \label{sec:opp-lines}}

\begin{figure}
\centerline{ \includegraphics[width=2.0 in]{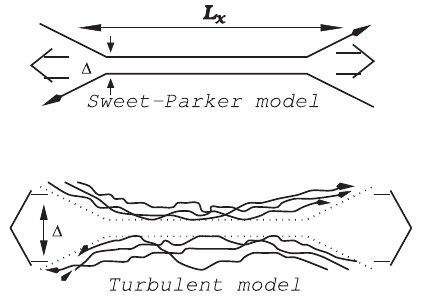}}
\caption{The relationship between Sweet-Parker and the view of turbulent reconnection of Lazarian et al., Phys. Plasmas \textbf{27}, 012305 (2020) is illustrated in their Figure 4.  In the Sweet-Parker model, the thickness of the outflow $\Delta$ is limited by Ohmic diffusivity, which makes reconnection slow.  In their turbulent model, $\Delta$ is determined by macroscopic field line wandering, which can be very wide.  In the modern reconnection literature, plasmoids are often invoked to widen $\Delta$. } 
\label{fig:recon}
\end{figure}

The Sweet-Parker model of magnetic reconnection,   which is the top illustration in Figure \ref{fig:recon}, remains the most common model of reconnection and is essentially the figure shown in papers on plasmoid models of reconnection \cite{Plasmoids,Plasmoid:2019}.   The whole of Figure \ref{fig:recon} is from the  Lazarian et al \cite{Turb-recon:2020} review.  The bottom illustration explains their view that turbulence can enhance the rate reconnection by broadening the width $\Delta$ of the outflow region.

Although it is easy to understand why Figure \ref{fig:recon} would arise in a paper trying to explain fast reconnection in two-dimensional systems, it is surprising that is is included in the Lazarian et al review.  On page 2 of their review they explain that reconnection is too ubiquitous to require special conditions to occur and that it is not only the speed of reconnection but also what triggers reconnection that must be explained.  On page 49, they recognized that two dimensional models are missing physics that is essential to understanding reconnection in a three dimensional world.

\color{black}

Despite the  Sweet-Parker model representing the common view, it is an unlikely explanation for the ubiquity of magnetic reconnection in a universe with three spatial dimensions.  The Sweet-Parker model:

(1)  Ignores magnetic field line chaos.  Faraday's Law can be written in a mathematically equivalent way Equation (\ref{B-ev}), which has the form of an advection-diffusion equation.  Whether the advective velocity is chaotic or not is the primary determinant of the nature of solutions to  advection-diffusion equations.  A two-coordinate magnetic field problem is exceptional for an advective-diffusion equation, for then and only then is an exponentially increasing force required to maintain the flow.  

(2) Ignores the virial theorem.     A collision between two regions of oppositely directed magnetic field that were initially separated into compact regions of space with little magnetic field between them is difficult to reconcile with  the 1953 virial theorem of Chandrasekhar and Fermi \cite{Virial:1953}, which is derived in Appendix \ref{sec:Virial}.  The Chandrasekhar-Fermi  virial theorem proves that autonomous regions of magnetic field cannot exist for a timescale long compared to an inertial, or equivalently Alfv\'en, disassembly time.  A longer lifetime than Alfv\'enic for a magnetic configuration implies additional forces: gravity in stars or rigid coils in laboratory experiments and a method for the magnetic forces to be transmitted to places where gravitational effects are strong or coils are located.

\color{black} 

(3) Ignores the way two tubes collide in three-dimensional space.  Reconnection is often described as a collision between two tubes of magnetic flux.  A simple model of tubes colliding is a collision between two drinking-straws.  Except for the case of perfect alignment, the two straws will come into contact at a point.  Even in the case of perfect alignment, their contact is along a line, not a surface.  Surface contact requires not only perfect alignment but also interlocking surface shapes.

(4) Ignores the smoothness of $\vec{B}$.  Two Maxwell equations involve spatial derivatives of $\vec{B}$; the spatial dependence must be continuous and smooth.  The thinner a layer over which a magnetic field reverses direction, the larger the current density must be.  Assuming reconnection occurs only when the current density becomes arbitrarily large is not a  compelling \color{black} explanation for the ubiquity of reconnection.

The opposing field structures of the Sweet-Parker model,  Figure \ref{fig:recon}, were invented to allow fast reconnection when $\vec{B}$ depends on only two coordinates.  But, how can naturally flowing plasmas in three dimensions produce such unlikely magnetic structures while avoiding making the magnetic field chaotic?  Magnetic field line chaos, which naturally arises even in an ideal evolution with three dimensional flows, quickly leads to reconnection no matter how small the resistivity. This is due to the exponential distortion of tubes of magnetic flux, Figures \ref{fig:tubes} and \ref{fig:dist}.


As discussed in Section  \ref{sec:power-req}, most of the energy released when magnetic field lines break generally goes into the non-dissipative term $\vec{u}_\bot\cdot \vec{f}_L$, Section \ref{sec:energy-ev}.  \color{black} For reconnection throughout a volume, \color{black} this means into Alfv\'en waves, which can only be damped when narrow and intense current or vorticity sheets are formed.   Lazarian \cite{Lazarian} has discussed the damping of Alfv\'en waves in turbulent plasmas, where the nature of the damping depends on the wavelength of the waves relative to the size of the turbulent eddies.   Even when the plasma is not turbulent, sheet currents quickly arise to damp Alfv\'en waves when the field lines are chaotic \cite{{Boozer:rec2023},Heyvaerts-Priest:1983,Similon:1989}, so the total time until their occurrence is $\sim 20\tau_{ef}$.  The currents sheets that arise to damp  Alfv\'en waves can be confused with the current sheets of the Sweet-Parker model.   Huang and Bhattacharjee observed a fast formation of current sheets in their important paper on reconnection driven by a chaotic flow \cite{Huang-Bhattacharjee}.  Their definition of reconnection, which was the damping of the released energy rather than the breaking of the connections of the field lines themselves, naturally leads to a confusion.


\section{Boundary conditions on $\vec{B}$ \label{sec:bndy} }

To rigorously deal with the breaking of magnetic field line connections, boundary conditions are needed in order to have connections that can be broken.  These can be dealt with by a perfectly conducting boundary---such as a sphere or the box of Figure \ref{fig:dist}---about the region that is to be studied.  External forcing is then represented by a movement of the boundary.  In toroidal plasmas, the two periodicities of the torus can act in a way similar to boundary conditions in defining broken connections.

A periodic box is sometimes assumed in simulations of reconnection.  \color{black} Assuming periodicity in a non-periodic system introduces unphysical points at which a field line closes on itself after transversing a number of periods.  The existence of such points follows from Brouwer's fixed-point theorem:  When any continuous function is mapped from a compact convex set to itself, there is a point $\vec{x}_0$ such that $\vec{f}(\vec{x}_0) = \vec{x}_0$.

Such fixed points are a central element in tearing mode theory in three dimensions.  \color{black} Tearing modes in tokamaks \color{black}  only arise on surfaces on which magnetic field lines close on themselves.  Plasmoids are generally viewed as arising from tearing modes \cite{Plasmoids}, but closed magnetic field lines seem unlikely in the extreme in three-dimensional naturally-occurring plasmas though they are present in two-dimensional models. \color{black} Such tearing modes must be given by a ballooning-mode formulation, such as that discussed in 2019 by Zhu et al. \cite{Ballooning-Tearing:2019}. \color{black}


\section{Large-scale flows with turbulence \label{Large-scale flows}}

Plasmas both natural and laboratory are almost always turbulent.   Nevertheless, important physical effects may be determined by the large-scale flow rather than the direct effects of the turbulent eddies associated with these flows.  \color{black} In the theory of chaotic dynamical systems such large scale flows are called L\'evy flights. \color{black}  Large-scale magnetic fields are probably more important in effects, such as the collapse of gas clouds in galaxies to form stars, than fields on the scale of the turbulent eddies.

The two highest Reynolds number fluids with which everyone is familiar are the oceans and the atmosphere.   In the atmosphere, large scale prevailing winds, such as the westerlies and easterlies, determine much of the weather and allowed sailing ships to reliably cross the Atlantic.  The large scale flow in the Atlantic itself, the Gulf Stream, has a profound effect on weather.   The weather in Britain would be different if the turbulent eddies determined the transport of tropical water to its shores.

The explanation of large-scale atmospheric and ocean flows is controversial, but their existence is not.  Concern about the effects of global warming on the flows underlies much of the research \cite{Westerlies}.  Zonal flows  \cite{Zonoflows} are one of these explanations, a theory that was largely developed for application to plasma microturbulence but also to Jupiter's zonal stripes by Diamond and collaborators \cite{Diamond}.  Where a plasma has a large scale flow, it is natural to assume the magnetic field does as well; the magnetic field and the plasma flows are related.

\begin{figure}
\centerline{ \includegraphics[width=1.5in]{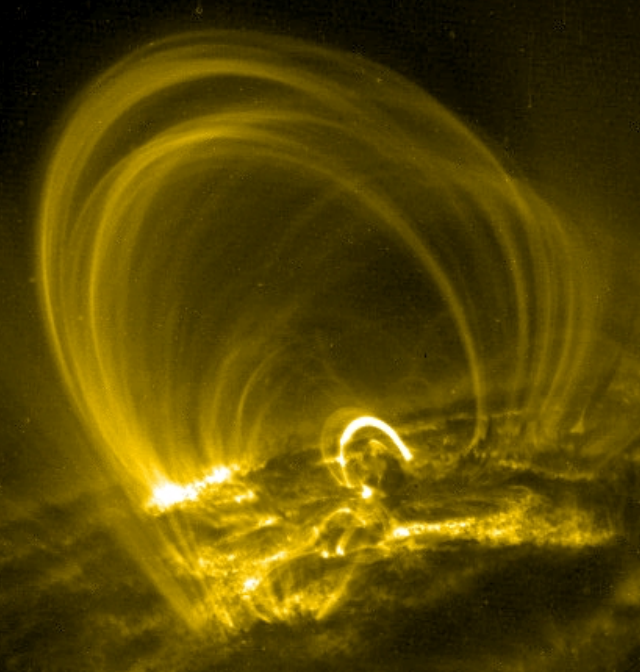}}
\caption{The \emph{Transition Region And Coronal Explorer} (TRACE) produced many ultra-violet images of the solar coronal loops during its 1998-2010 mission.  The one illustrated is T171\_20050908\_114211X17.  The bright lines are generally thought to trace magnetic field lines, which have length to width ratios $L/a_\bot\sim40$. } 
\label{fig:corona}
\end{figure}

It is thought that coronal loops, Figure \ref{fig:corona}, represent large scale magnetic fields which evolve due to footpoint plasma motion, which is presumably turbulent.  The field evolution given in Figure \ref{fig:dist}  can be interpreted as a simple model for coronal loops.   Although the coronal loops are thought to be regions of magnetic field concentration, when the width of the current profile that produces the loop is comparable to its width, the resistive timescale $\mu_0 a_\bot^2/\eta$ is far longer than the lifetime of the loop.

As discussed in Section \ref{sec:power-req} and in Section II.D of \cite{Boozer-null-X}, a length/width ratio of approximately $\ln(\mathcal{R}_a)$ is to be expected for loops that are produced by chaotic flows.  The reason is simple.  The distance a flow of speed $v$ can cover before becoming incoherent because of diffusion is $L\approx \big(\tau_{ef}\ln(\mathcal{R}_a\big)v$, but $\tau_{ef}\approx a_\bot/v$, since the speed at which streamlines can exponentiate apart is limited by the gradient of the flow across the stream lines.  Consequently, $L/a_\bot\approx\ln(\mathcal{R}_a)$.  A critical $L/a_\bot$ ratio for reconnection is consistent with the coronal mass ejections discussed by Gou et al. \cite{Gou}.   Ultra-violet images of coronal loops  suggest length to width ratios of approximately forty in what are presumptively chaotic magnetic fields, Figure \ref{fig:corona}.  \color{black}

When it is assumed that turbulent eddies dominate the flow, even at the largest spatial scales, mixing can be greatly slowed.  \color{black} With the same maximum flow speed, mixing due to large scale chaotic flows is far faster than with small scale chaotic flows.  \color{black}  The review of fluid turbulence by Falkovich, Gaw\c{e}dzki, and Vergassola \cite{Turb-Mixing} dealt with the separation of fluid elements by turbulent eddies, rather than large scale flows.  The turbulent separation obeys a power law in time, not exponential.  For this reason, the existence of large scale flows is essential for understanding the speed of mixing over long scales.    This result is well known to anyone who has stirred soup in a wide but shallow pot.  What is important is to move the soup side to side on the scale of the pot.  The creation of many highly localized eddies by a complicated stirring mechanism is not as efficient. 

\section{Discussion \label{sec:discussion}}   

The success of a research in building a physics understanding of magnetic reconnection and dynamos depends on the validity of the fundamental assumptions.   The theory of magnetic reconnection developed out of two-dimensional models.  As explained in Section \ref{2D-recon}, the physics of two-dimensional reconnection has remarkably little relevance to reconnection in three dimensions, which resulted in research being based on a number of false assumptions:

(1)  Neglect of magnetic field line chaos.   In two dimensions, a chaotic field-line flow is precluded because it causes an exponential increase in  the magnetic energy.  Flows naturally move in the direction of least resistance (meaning back force not electrical resistance).   As shown in  Equation (\ref{f_L inequality}) and the related discussion, in three dimensions, a chaotic field line flow naturally flows in the direction in which the increase in the magnetic energy is only moderate even as the field lines become highly chaotic.  

Magnetic field line chaos at a certain level makes reconnection unavoidable, which explains why reconnection is so ubiquitous and places interesting constraints on the time and geometry requirements for reconnection.  \color{black} The natural tendency of magnetic field lines to be chaotic in three dimensional systems is painfully well known to anyone who has built a stellarator plasma confinement device. \color{black}

With chaos the power dissipated by a magnetic reconnection is negligible compared to the energy released by the breaking of field line connections.  The implication is that the released energy must go into \color{black} plasma motion involving the Alfv\'en speed, such as \color{black} Alfv\'en waves.  In a chaotic region, shear Alfv\'en waves evolve on an Alfv\'enic timescale  to form current sheets that cause a rapid dissipation \cite{Boozer:rec2023}.  Although these current sheets are distinct from those of two-dimensional reconnection theory, defining reconnection as the fast dissipation of magnetic energy rather than the breaking of field line connections confuses the physics.

\color{black} When chaos is neglected, the current density required for reconnection and for resistively dissipating the energy released by breaking the magnetic field line connections are comparable, $j\sim v B/\eta$, where $v$ is the evolution speed of the plasma and field.  With chaos, the required current density for producing reconnection is only $j\sim B/\mu_0 \ell_{ef}$, which is smaller by a factor of $\mathcal{R}_a$ of Equation (\ref{R_a}), often called the magnetic Reynolds number.  But, the required current density for resistively dissipating the energy released by the reconnection remains $j\sim v B/\eta$. \color{black}

(2) Neglect of magnetic helicity conservation.  Although helicity is dissipated by resistivity through the term $2\int \vec{B}\cdot\vec{E}d^3x$, its dissipation is extremely slow compared to the dissipation of magnetic energy, $\int \vec{j}\cdot\vec{E}d^3x$.  Any non-smoothness in the spatial dependence of the magnetic field increases the speed of energy dissipation compared to helicity dissipation.  The more turbulent the plasma the better helicity is conserved relative to magnetic energy.  The magnetic energy quickly relaxes to Woltjer's  minimum value \cite{Woltjer:1958} that is consistent with its helicity content.   

Magnetic dynamos would be easier to explain if turbulence created helicity.  Turbulence cannot without placing an enormous drag force on the plasma flow for a given rate of helicity creation---the shorter the spatial scale of the turbulence the stronger is this effect.  Helicity conservation does not preclude dynamos since they can be produced by helicity transport. 

\color{black} The accumulation of helicity due to its \color{black} conservation provides an explanation for coronal mass ejections.  Twisting of footpoints of coronal loops adds helicity that cannot be dissipated or removed by reconnection.  When the helicity becomes sufficiently large, the associated magnetic energy and force become too great to be coupled to regions dominated by gravity, and ejection is the natural consequence. 

\vspace{0.2in}

Not all faulty assumptions come from intuition based on two-dimensional analysis.  Three common assumptions are also false in two-dimensional reconnection:

(1) Neglect of the distinction between the plasma and the field line velocity.  This distinction is required to demonstrate that reconnection is given by $\mathcal{E}\neq0$, where $\mathcal{E}$ is the difference between $E_{||}$ and $-\partial\Phi(\vec{x})/\partial \ell$ with $\Phi$ a single-valued potential and $\ell$ the distance along a field line.  The plasma velocity $\vec{v}$ is directly affected by $\eta_\bot \vec{j}_\bot$ and the Hall term in Ohm's law, but the magnetic field line velocity $\vec{u}_\bot$ is not.

(2)  Neglect of the virial theorem.  The virial theorem derived by Chandrasekhar and Fermi \cite{Virial:1953} shows that magnetic field configurations that are not adequately coupled to regions with strong gravitational attraction or rigid coils will disassemble on an Alfv\'enic timescale. 

(3)  Neglect of the point-like intersection of general flux tubes.  When two flux tubes first collide they do so only at points unless they have two symmetry directions. 

\hspace{0.2in}

When understood, simple physics and mathematics considerations clarify magnetic reconnection and dynamos over a broad range of natural and laboratory plasmas---whether turbulent or not. With these insights, numerical simulations could greatly enhance our understanding.  The prejudices acquired from the theory of magnetic reconnection in two-dimensional systems need to be overcome while studying problems in which all three spatial coordinates are of importance.

The physics arguments in this paper are of continuing importance but are often ignored rather than disputed.  A major 2023 review of solar dynamos by Charbonneau and Sokoloff \cite{solar dynamo: 2023} invokes resistive helicity dissipation at small scales to restore the $\alpha$ effect but did not explain how small scale flows would invalidate helicity conservation.  The 2025 article by Muraglia, Agullo, Dubuit, Bigu\'e, and Garbet \cite{Tok-recon:2025} on reconnection in toroidal plasmas emphasized resistive tearing modes  and ignored chaos.  Jardin et al. \cite{Jardin-fast:2022} found that chaos allowed ideal instabilities that do not self-saturate to result in an extremely fast reconnection that is independent of the plasma resistivity.

\color{black}

\section*{Acknowledgements}

This material is based upon work supported by the U.S. Department of Energy, Office of Science, Office of Fusion Energy Sciences under Award DE-FG02-95ER54333.  The numerous questions of the referees helped me make this importance of this paper clearer. \\

\section*{Author Declarations}

The author has no conflicts to disclose. \vspace{0.01in}


\section*{Data availability statement}

Data sharing is not applicable to this article as no new data were created or analyzed in this study.


\appendix
\section{Virial Theorem \label{sec:Virial}}

The virial theorem is a general constraint on force balance in mechanical systems.  The virial theorem including magnetic fields  was derived in 1953 by
 S. Chandrasekhar and E. Fermi \cite{Virial:1953}.  It is rarely derived or even mentioned in plasma texts, an exception is \color{black} page 72 of G. Schmidt's book \cite{Virial}.

Let $d\vec{x}/dt = \vec{v}(x,t)$ be the mass flow velocity of a plasma.  The moment of inertia of a plasma within a region of space is
\begin{eqnarray}
I &\equiv& \int \rho_m |\vec{x}|^2 d^3x \\
&=& \sum_i m_i |\vec{x}_i|^2,
\end{eqnarray}
when the volume is divided into many cells each containing a mass of plasma $m_i$.  The second time derivative of the moment of inertia is
\begin{eqnarray}
\frac{d^2I}{dt^2} &=& 2 \sum_i m_i |\vec{v}_i|^2 + 2 \sum_i m_i \vec{x}_i \cdot \frac{d\vec{v}_i}{dt} \\
&=& 2 \int \rho_m |\vec{v}|^2 d^3x + 2\int \vec{x}\cdot \vec{F} d^3x, \mbox{    where   } \hspace{0.2in}\\
\rho \frac{d\vec{v}}{dt} &=& \vec{F}.
\end{eqnarray}
The force exerted on a plasma is
\begin{eqnarray}
\vec{F} &=& - \vec{\nabla}\cdot (\tensor{p} + \tensor{T}) +\vec{F}_a;\\
\tensor{T} &=& \frac{B^2}{2\mu} \tensor{1} - \frac{\vec{B} \vec{B}}{\mu_0} \mbox{    where   } \\
\vec{\nabla}\cdot\tensor{T} &=& \frac{\vec{\nabla}\times\vec{B}}{\mu_0} \times \vec{B}.
\end{eqnarray}
$\vec{F}_a$ is any additional force that is not thermal or magnetic; $\tensor{p}$ is the plasma pressure tensor, which includes viscous forces.
\begin{eqnarray}
\int \vec{x} \cdot (\vec{\nabla}\cdot (\tensor{p} + \tensor{T})) d^3x &=&- \int \tensor{1}:(\tensor{p} + \tensor{T})d^3x \nonumber \\
&&  +\int \vec{\nabla}\cdot(\vec{x}\cdot(\tensor{p} + \tensor{T}) d^3x \hspace{0.11in} \nonumber\\
&=& - 3p - \frac{B^2}{2\mu_0} \nonumber\\ &&+ \oint \vec{x}\cdot(\tensor{p}  + \tensor{T}))\cdot d\vec{a}. \hspace{0.1in}
\end{eqnarray}

\begin{eqnarray}
\frac{1}{2}\frac{d^2I}{dt^2} &=& 2U_{KE} + 2 U_{th} + U_B +\int\vec{x}\cdot\vec{F}_a d^3x \nonumber\\
&& - 2\oint \vec{x}\cdot(\tensor{p}  + \tensor{T}))\cdot d\vec{a}; \\
U_{KE} &\equiv& \frac{1}{2} \int \rho |\vec{v}|^2 d^3x;\\
U_{th} &\equiv& \frac{3}{2} \int p d^3x;\\
U_B &\equiv& \frac{1}{2\mu_0} \int B^2 d^3x.
\end{eqnarray}

An autonomous magnetic field structure can be surrounded by a surface on which the field is zero, which implies
\begin{equation}
\frac{1}{2}\frac{d^2I}{dt^2} = 2U_{KE} + U_B.
\end{equation}
Since the terms on the right-hand side of this equation are positive, the moment of inertia satisfies $d^2I/dt^2>0$, and the magnetic structure must disassemble on an Alfv\'enic timescale.


\color{black}


\end{document}